\documentclass[useAMS,usenatbib]{mn2e}
\usepackage{epsfig}
\usepackage{graphics}
\usepackage[usenames]{color}
\usepackage{amssymb}
\usepackage{amsmath}
\usepackage{times}
\newcommand{\be}{\begin{equation}}
\newcommand{\ee}{\end{equation}}
\newcommand{\bdm}{\begin{displaymath}}
\newcommand{\edm}{\end{displaymath}}
\newcommand{\bea}{\begin{eqnarray}}
\newcommand{\eea}{\end{eqnarray}}

\newcommand{\sesa}[1]{\textcolor{magenta}   {\texttt{\textbf{SESA: #1}}} }

\newcommand{\msun}{M_\odot}
\def\lsim{\lower.5ex\hbox{$\; \buildrel < \over \sim \;$}}

\title[Evolution of massive black hole binaries]
{Scattering experiments meet N-body I: a practical recipe for the evolution of massive black hole binaries in stellar environments}

\author[A. Sesana and F. M. Khan and ]{Alberto Sesana$^{1,2}$\thanks{E-mail: asesana@star.sr.bham.ac.uk} and Fazeel Mahmood Khan$^{3}$\thanks{E-mail:khanfazeel.ist@gmail.com}\\
$^{1}$ School of Physics and Astronomy, University of
Birmingham, Edgbaston, Birmingham B15 2TT, United Kingdom \\
$^{2}$ Max-Planck-Institut f{\"u}r Gravitationsphysik, Albert Einstein
Institut, Am M{\"u}hlenberg 1, 14476 Golm, Germany \\
$^{3}$ Department of Space Science, Institute of Space Technology, P.O. Box 2750 Islamabad, Pakistan\\
}

\begin{document}

\date{}

\pagerange{\pageref{firstpage}--\pageref{lastpage}} \pubyear{2010}

\maketitle

\label{firstpage}

\begin{abstract}
  The N-independence observed in the evolution of massive black hole binaries (MBHBs) in recent simulation of merging stellar bulges suggests a simple interpretation beyond complex time-dependent relaxation processes. We conjecture that the MBHB hardening rate is equivalent to that of a binary immersed in a field of unbound stars with density $\rho$ and typical velocity $\sigma$, provided that $\rho$ and $\sigma$ are the stellar density and the velocity dispersion at the influence radius of the MBHB. By comparing direct N-body simulations to an hybrid model based on 3-body scattering experiments, we verify this hypothesis: when normalized to the stellar density and velocity dispersion at the binary influence radius, the N-body MBHB hardening rate approximately matches that predicted by 3-body scatterings in the investigated cases. The eccentricity evolution obtained with the two techniques is also in reasonable agreement. This result is particularly practical because it allows to estimate the lifetime of MBHBs forming in dry mergers based solely on the stellar density profile of the host galaxy. We briefly discuss some implications of our finding for the gravitational wave signal observable by pulsar timing arrays and for the expected population of MBHBs lurking in massive ellipticals. 


\end{abstract}


\section{Introduction}
\label{sec:Introduction}
Massive black holes (MBHs) are fundamental building blocks in the process of galaxy formation and evolution; they are ubiquitous in nearby galaxy nuclei \citep[see, e.g., ][]{mago98}, and their masses correlate with the properties of the host bulge \citep[][and reference therein]{kormendy13}. If MBHs are common in galaxy centres at all epochs, as implied by the fact that galaxies harbour active nuclei for a short period of their lifetime \citep{hr93}, then a natural consequence of the hierarchical paradigm of structure formation is that a large number of massive black hole binaries (MBHBs) forms along the cosmic history, following galaxy mergers \citep{bbr80}.

As galaxies build-up, they keep turning cold gas into stars, with the result that massive galaxies at low redshift are predominantly gas poor. Nonetheless, massive gas poor galaxies continue to merge with satellites in groups and clusters \citep[see, e.g.][]{vandokkum05}. In fact, a significant fraction of them is observed in close pairs \citep[see, e.g.][]{lopez12,xu12}, implying a significant merger rate. The fate of the forming MBHB in these massive, gas poor galaxy mergers have attracted the attention of several investigators that tackled the problem either (semi)analytically \citep[e.g.][]{quinlan96,yu02,sesana07}, or by means of direct N-body simulations \citep[e.g.]{mm01,makino04,berczik06}.

In recent years, advancement in massive parallel computing made possible to simulate 'ab initio' the evolution of MBHBs in merging bulges, using up to a few million particles \citep{preto11,khan11,khan12,gualandris12}. These authors found that the MBHB hardening rate (the rate at which its semimajor axis shrinks) is independent of the number of particles, suggesting efficient diffusion of stars in phase space, keeping the binary 'loss cone' -- i.e. the portion of stellar distribution phase space collecting orbits with small angular momentum, intersecting the binary semimajor axis -- full.

In this letter we shall assume that these simulations capture the relevant physics governing the diffusion of stars in phase space in the merger remnant, and we interpret this in light of simple 3-body scattering theory. By comparing direct N-body simulations to an hybrid model based on 3-body scattering experiments \citep[developed by][]{sesana10}, we show that, when normalized to the stellar density and velocity dispersion at the binary influence radius, the N-body MBHB hardening rate  matches that predicted by 3-body scattering. This result is extremely practical, because the MBHB coalescence time can then be reliably estimated once the density profile of the stellar distribution of the host galaxy is known. 

The paper is organized as follows. In Section 2, we describe the N-body runs, the hybrid model and the match between the two. We present our results in Section 3 and we discuss its importance and possible applications in Section 4.

\section {Matching N-body simulations to the hybrid model}
\label{sec2}

\subsection{N-body runs}
\label{sec2.1}
The N-body runs performed for this study, closely follow those described in \citep[][hereinafter K12]{khan12}. The two merging bulges are initialized as equal mass and size Dehnen profiles \citep{dehnen93}:
\begin{equation}
\rho(r)=\frac{(3-\gamma)M_*}{4\pi}\frac{r_0}{r^{\gamma}(r+r_0)^{4-\gamma}},
\label{rhoD}
\end{equation}
where $M_*$ is the total mass of the stellar bulge, $r_0$ is the scale radius, and $\gamma$ is the inner logarithmic slope. A MBH is placed at the centre of each bulge. We perform four simulations, by assuming two different inner cusp slopes -- $\gamma=1,1.5$ -- and MBHB mass ratios -- $q=M_2/M_1=1,1/3$, where $M_1>M_2$ are the masses of the primary and secondary MBH respectively. In all cases we set $M_1=0.005 M_{\rm Gal}$. The two bulges are initially at a separation 15 on an bound orbit with eccentricity 0.75. The units adopted for the integration are $G=M_*=r_0=1$. $N$-body simulations are performed using $\phi-$GRAPE+GPU on \textit{accre}, a high-performance GPU computing cluster at Vanderbilt University, Nashville, TN. The code is updated version of \cite{2007NewA...12..357H} and is described in section 2.2 of \cite{2013ApJ...773..100K}.

\subsection{Hybrid model}
\label{sec2.2}
The hybrid model we use has been extensively described in \citep[][hereinafter S10]{sesana10}. For a given density profile -- e.g. the one given in equation (\ref{rhoD}) --, one can compute the initial separation $a_0$ at which the mass enclosed in the binary is twice the mass of the secondary (i.e. $M_*(<a_0)=2M_2$). At this point, an initial eccentricity $e_0$ is also assumed, and the binary is evolved forward in time. The evolution takes into account for the initial scattering of bound stars leading to the erosion of the central stellar cusp \citep[see full details in][]{shm08}, followed by a phase dominated by scattering of unbound stars intersecting the binary semimajor axis \citep{quinlan96,sesana06}, and the efficient gravitational wave (GW) emission stage \citep{pm63} leading to final coalescence of the system. After the short phase of cusp disruption, in the unbound scattering phase, the binary hardening proceeds at a rate

\begin{equation}
\frac{d}{dt}\left(\frac{1}{a}\right)=\frac{G\rho}{\sigma}H_{\rm 3b},
\label{aev}
\end{equation}
where $H_{\rm 3b}\approx15-20$ is a dimensionless rate, and $\rho$ and $\sigma$ are effectively free parameters that, in the original formulation of the 3-body scattering problem, represent the density and velocity of the distribution of intruding stars at infinity. The fact that $H_{\rm 3b}$ is found to be independent of $a$ (for hard binaries) means that the hardening rate given by equation (\ref{aev}) is about constant in time. In the framework of the hybrid model, one can tune the supply of unbound stars to the binary by picking some specific value of $\rho$ and $\sigma$ related to the adopted stellar distribution -- equation (\ref{rhoD}). Units are set so that $M=M_1+M_2=G=a_0=1$ and all the technical details can be found in S10.

\subsection{Comparing the MBHB evolution in the two cases}
\label{sec2.3}
As shown in the previous subsections, the N-body and the hybrid approaches have been developed using different units, which makes a comparison between the two somewhat cumbersome. One can assume the density profile given by equation (\ref{rhoD}) and then convert the units of one model into the other, but this is not the best way to proceed, because the initial density profile of the N-body model evolves during the merger, and the stellar distribution in which the binary forms and evolves {\it is different} from the initial one. In some sense, we want to {\it normalize the binary evolution to the stellar environment}. Because of the fairly constant hardening rate seen in N-body simulations, it is natural to also write it in the form given by equation (\ref{aev}), where one has the freedom to pick an arbitrary value of $\sigma$ and $\rho$, which results in a different value of $H_{\rm Nb}$ \citep[this approach has already been employed by][]{khan11,gualandris12}. 

We conjecture that the $N$-body MBHB hardening rate is equivalent to that of a binary immersed in a field of unbound stars with density $\rho_{\rm inf}$ and velocity $\sigma_{\rm inf}$, equal to the density and velocity dispersion of the surrounding stellar distribution {\it at the binary influence radius}. The latter is defined as the radius ,$r_{\rm inf}$, containing twice the binary mass in stars -- i.e., $M_*(<r_{\rm inf})=2M$. This statement is equivalent to say that the binary loss cone is full at $r_{\rm inf}$ (see discussion in Section 3.1 in S10). The explanation of why the loss cone is full is beyond the scope of this letter, and it must trace back to efficient stellar diffusion in the time-dependent, triaxial, rotating potential of a merger remnant. If our conjecture is correct then by plugging the $(d/dt)(1/a)$ measured in the N-body simulation, $\rho_{\rm inf}$ and $\sigma_{\rm inf}$ into equation (\ref{aev}), we should obtain  $H_{\rm Nb}\approx H_{\rm 3b}$. This also means that the evolution of the MBHB in the two approaches should be remarkably similar if the same time units are used. Here, for 'same time units' we mean units that normalize the stellar density and velocity in equation (\ref{aev}) to the same value in the N-body run and in the hybrid model. In practice the N-body and hybrid codes have their own units in which:
\begin{equation}
\frac{d}{dt_{\rm 3b}}\left(\frac{1}{a_{\rm 3b}}\right)=\frac{G\rho_{\rm 3b}}{\sigma_{\rm 3b}}H_{\rm 3b},
\label{aev}
\end{equation}
\begin{equation}
\frac{d}{d_{\rm Nb}t}\left(\frac{1}{a_{\rm Nb}}\right)=\frac{G\rho_{\rm Nb}}{\sigma_{\rm Nb}}H_{\rm Nb}.
\label{aev}
\end{equation}
Therefore, assuming as a reference the N-body time, the MBHB evolution obtained in the hybrid formalism will match that of the N-body runs if the time in the hybrid model is rescaled according to:
\begin{equation}
t_{\rm Nb}=\frac{a_{\rm Nb}}{a_{\rm 3b}}\frac{\sigma_{\rm Nb}}{\sigma_{\rm 3b}}\frac{\rho_{\rm 3b}}{\rho_{\rm Nb}}t_{\rm 3b},
\label{tnorm}
\end{equation}
where $\sigma_{\rm Nb}=\sigma_{\rm inf}$ and $\rho_{\rm Nb}=\rho_{\rm inf}$, and all quantities are measured in the respective units.  

\section{Results}
\label{sec3}

\begin{figure*}
\begin{tabular}{cc}
\includegraphics[scale=0.40,clip=true]{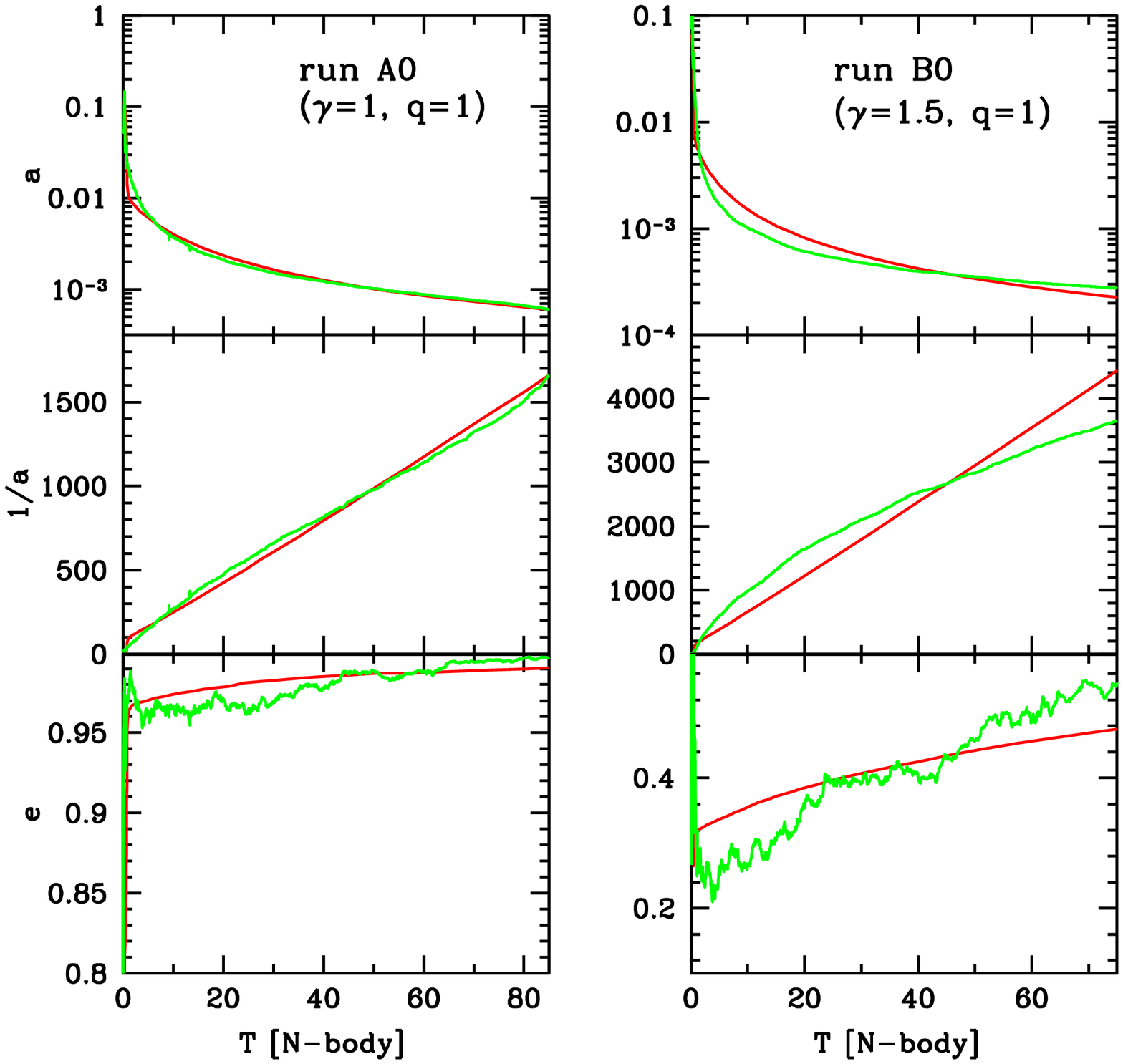}&
\includegraphics[scale=0.40,clip=true]{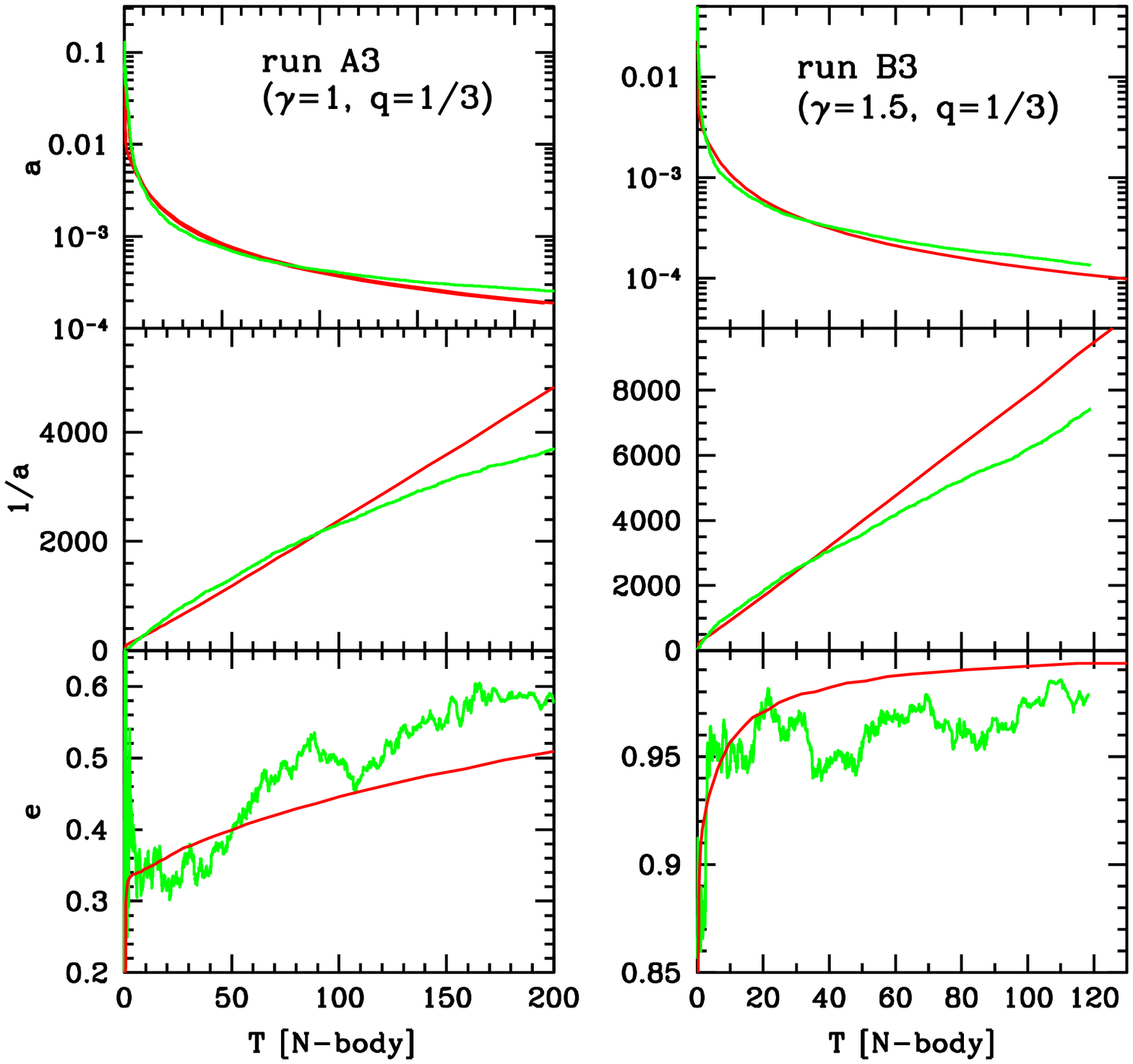}\\
\end{tabular}
\caption{MBHB evolution in the N-body runs and in the hybrid model normalized to the N-body units. $T=0$ has been set to the moment when the binary reaches the separation $a=0$. In each column of plots we show (from the top to the bottom) the time evolution of the semimajor axis, the inverse of the semimajor axis and the eccentricity. In each panel, the green curve is the result of the N-body simulation, the red curve is the evolution predicted by the hybrid model with the same parameters.}
\label{fig1}
\end{figure*}

\subsection{Hardening rates}
\label{sec3.1}
A comparison between the four N-body runs and the hybrid models featuring the same parameters is shown in figure \ref{fig1}. The evolution is rescaled to N-body units, i.e. the time unit in the hybrid model has been converted according to equation (\ref{tnorm}). Moreover the time has been shifted so that $T=0$ when the MBHB sits at a semimajor axis $a_0$. We recall that $a_0$ is defined so that $M_*(<a_0)=2M_2$, and it is the starting point of the hybrid model. We see in the top panels that, in all cases, the evolution of the semimajor axis matches quite well in the two approaches, as expected. Differences between the N-body runs and the hybrid model become more visible when $1/a$ is plotted (central panels). Here we see that we still have a good match, even though in the N-body runs a somewhat slower evolution is visible. A slight decline of the shrinking rate can be due to either (i) a not completely full loss cone or (ii) a declining density of the underlying stellar distribution at the influence radius. 

\begin{figure}
\includegraphics[scale=0.55,clip=true,angle=270]{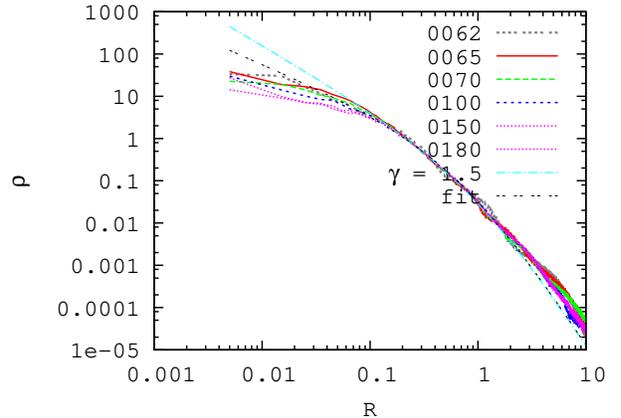}
\caption{Density profile of simulation B3 at different times (see labels in figure). The cyan dash-dotted line is the density profile of the initial model ($\gamma_i=1.5$ and the black double-dashed line is the best fit to the density profile when the binary becomes bound ($\gamma=1.1$).}
\label{fig2}
\end{figure}

To gain insights on this point, we plot in figure \ref{fig2} the stellar density profile of the B3 N-body run at different snapshot (we should keep in mind that in figure \ref{fig1} we normalized $T$ to the moment when the binary becomes bound, which corresponds to snapshot 0062 in figure \ref{fig2}). In run B3, $r_{\rm inf}$ is located at 0.084 (see table \ref{tab1}), and we notice that at this separation the stellar density declines constantly with time. The drop ranges between 10\% and 40\% depending on the simulation, naturally slowing down the MBHB evolution by an amount comparable to what seen in figure \ref{fig1}. 

\begin{table}
\begin{center}
\begin{tabular}{ccccccc}
\hline
Model & $\gamma_i$ & $q$ & $r_i$ & $a_0$ & $H_{\rm Nb}$ & $H_{\rm 3b}$ \\
\hline
A0 & 1.0 & 1   & 0.16  & 0.113  & 14.53 & 17.4\\
A3 & 1.0 & 1/3 & 0.13  & 0.066  & 14.29 & 16.4\\
B0 & 1.5 & 1   & 0.10  & 0.066  & 11.43 & 16.5\\
B3 & 1.5 & 1/3 & 0.084 & 0.039  & 12.46 & 17.7\\

\hline
\end{tabular}
\end{center}
\caption{Hardening rates for the 4 simulated MBHB systems (column 1). We report the initial inner slope of the density profile (column 2), the binary mass ratio (column 3), the binary influence radius and initial separation of the hybrid model in N-body units (columns 4 and 5), and the hardening rates estimated from the N-body simulations and from the hybrid model (columns 6 and 7).}
\label{tab1}
\end{table}

The hardening rate comparison is shown in table \ref{tab1}, where $H_{\rm Nb}$ has been computed by fixing $r_{\rm inf}$ at the time the MBHB first becomes bound (column 2 in the table) and averaging over the entire subsequent evolution in hard binary phase. As expected, we find $H_{\rm Nb}\approx H_{\rm 3b}$ (within 30\%), confirming our conjecture. We notice, however, that $H_{\rm Nb}$ is systematically lower, likely due to the evolution of the density profile discussed above. In fact, in computing $H_{\rm Nb}$ we fixed $r_{\rm inf}$ at $T=0$. However the subsequent density decline (figure \ref{fig2}), also implies a sizeable expansion of $r_{\rm inf}$ during the simulation. In practice $r_{\rm inf}$ is an evolving quantity, and any computation of a time averaged N-body hardening rate is intrinsically approximate. Therefore, fixing $r_{\rm inf}(T=0)$ is expected to return an average $H_{\rm Nb}$ that is lower to the one found in 3-body scatterings. 

Another remarkable aspect of this comparison is the quite similar evolution of the binary eccentricity in the two approaches (lower panels in figure \ref{fig1}), even though the N-body data are inevitably noisy. We caution, however, against some of the limitations of the hybrid model. One major caveat is that the bound cusp is unaffected by the presence of the binary until $M_2$ reaches $a_0$. Although this is a fair assumption for binaries with $q<1/10$, in the comparable mass cases investigated here $M_2$ will affect the distribution of stars around $M_1$ earlier in its evolution, implying a smoother transition between dynamical friction and scattering of stars refilling the loss cone. As a consequence, the binary evolution in the short initial cusp erosion phase in the hybrid model is in general faster than in the N-body simulation. This is not a major issue, since the binding energy transferred to the binary from the cusp erosion is always the same, whether this happens promptly or gradually. However, the angular momentum exchange can be significantly different, meaning that the eccentricity evolution predicted by the hybrid model might not be fully trustworthy in this phase. 

\subsection{Coalescence times}
Although we plan to address extensively this topic in a forthcoming paper, the results obtained in the previous section can be used to get an approximate estimate of typical lifetimes and characteristic separations of MBHBs forming in dry galaxy mergers. The evolution of the semimajor axis can be written as
\begin{equation}
\frac{da}{dt}=\frac{da}{dt}\Big{|}_{3b}+\frac{da}{dt}\Big{|}_{\rm gw}=-Aa^2-\frac{B}{a^3},
\label{aevtot}
\end{equation}
where
\begin{equation}
  A=\frac{GH\rho_{\rm inf}}{\sigma_{\rm inf}},  \,\,\,\, B=\frac{64G^3M_1M_2MF(e)}{5c^5}, 
\end{equation}
and $F(e)=(1-e^2)^{-7/2}[1+(73/24)e^2+(37/96)e^4]$ \citep{pm63}.
Since the stellar hardening is $\propto a^2$ and the GW hardening is $\propto a^{-3}$, binaries spend most of their time at the transition separation obtained by imposing $(da/dt)_{3b}=(da/dt)_{\rm gw}$:
\begin{equation}
  a_{*/{\rm gw}}=\left[\frac{64G^2\sigma_{\rm inf}M_1M_2MF(e)}{5c^5H\rho_{\rm inf}}\right]^{1/5},
\label{atrans}
\end{equation}
and their lifetime can be written as
\begin{equation}
  t(a_{*/{\rm gw}})=\frac{\sigma_{\rm inf}}{GH\rho_{\rm inf}a_{*/{\rm gw}}}.
\label{ttrans}
\end{equation}
In fact the lifetime estimated through equation (\ref{ttrans}) differs by only about 10\% from that obtained by integrating equation (\ref{aevtot}). To get physical estimates for realistic galaxies we need to estimate $\sigma_{\rm inf}$ and $\rho_{\rm inf}$. We get the former from the $M-\sigma$ relation as reported by \cite{kormendy13}, $M_9=0.309\sigma_{200}^{4.38}$, where $M_9$ is the binary mass normalized to $10^9\msun$ and $\sigma_{200}$ is the bulge velocity dispersion normalized to 200 km s$^{-1}$. The latter is obtained from the density profile given by equation (\ref{rhoD}) evaluated at the influence radius $r_{\rm inf}=r_0/\{[M_*/(2M)]^{1/(3-\gamma)}-1\}$, where $r_0$ and $M_*$ need to be specified. For a given MBHB mass $M$, we get $M_*$ from the $M-bulge$ relation as reported by \cite{kormendy13}, $M_9=0.49M_{*,11}^{1.16}$, where where $M_{*,11}$ is the bulge mass normalized to $10^{11}\msun$. The break radius $r_0$ is connected to the bulge effective radius $R_{\rm eff}$ through the relation $R_{\rm eff}\approx 0.75r_0(2^{1/(3-\gamma)}-1)^{-1}$ \citep{dehnen93}. This latter is a function of the galaxy stellar mass which depends on the nature of the galaxy host. In particular \cite{2008MNRAS.386..864D} found that $R_{\rm eff}/{\rm pc}={\rm max}(2.95 M_{*,6}^{0.596},34.8 M_{*,6}^{0.399})$ for elliptical galaxies, whereas  $R_{\rm eff}/{\rm pc}=2.95 M_{*,6}^{0.596}$ for the bulges of spirals and ultra compact dwarfs ($M_{*,6}$ is the mass of the stellar bulge normalized to $10^6\msun$).

\begin{figure}
\includegraphics[scale=0.42,clip=true,angle=0]{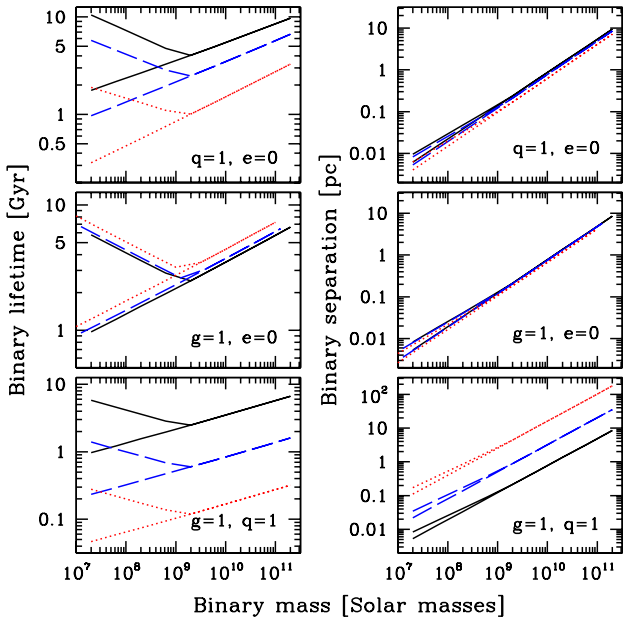}
\caption{Lifetime (left panels) and characteristic separation (right panels) of MBHBs vs binary mass. In the top panels solid--black, dashed--blue and dotted--red lines are for density profiles with $\gamma=0.5,1,1.5$ respectively; in the central panels solid--black, dashed--blue and dotted--red lines are for MBHBs with $q=1,0.32,0.1$ respectively; in the bottom panels solid--black, dashed--blue and dotted--red lines are for MBHB with $e=0,0.9,0.99$ respectively. In each bifurcated curve, the upper branch is for regular ellipticals, the lower branch is for bulges of spirals and ultra compact dwarfs.}
\label{fig3}
\end{figure}

With the relations described above, for each MBHB mass, we can estimate $M_*$, $r_0$, $r_{\rm inf}$, $\rho_{\rm inf}$, $\sigma_{\rm inf}$ of the host galaxy, and compute its lifetime and characteristic separation through equations (\ref{atrans}) and (\ref{ttrans}). Results are shown in figure \ref{fig3} for a range of MBHB and host galaxy properties. One of the consequences of our findings is that MBHBs in purely stellar environment do not stall, but their lifetimes are rather long, ranging from 0.1 to several Gyr, depending on the properties of the systems (left panels in figure \ref{fig3}). In particular, binaries with $M=10^9-10^{10}\msun$ hosted in giant ellipticals with shallow density profiles, which should dominate the GW signal in the nHz frequency regime accessible to pulsar timing arrays (PTAs) \citep{2008MNRAS.390..192S}, might have lifetimes as long as 2-10 Gyr, regardless of their mass ratio. These lifetimes, however, can be significantly shorter if eccentricity is driven to very high values during the stellar dominated evolution phase (as some of the tracks shown in figure \ref{fig1} might suggest). For example, for $e=0.99$, typical coalescence times are a factor of $\approx20$ shorter (lower left panel in figure \ref{fig3}). Regardless of their mass ratio or of the galaxy density profile most of this systems spend the majority of their time at a separation which is roughly proportional to the MBHB mass, and for $M=10^9-10^{10}\msun$ systems is in the range 0.1-1 pc. High eccentricity promotes efficient GW emission at larger separations, and binaries with $e=0.99$ besides having a lifetime which is $\approx20$ shorter, have a better chance to be found at separations which are $\approx10$ times larger (lower right panel in figure \ref{fig3}). 

\section{Discussion and conclusion}
We performed the first detailed comparison between direct N-body simulations of MBHB mergers in stellar bulges and the hybrid model based on 3-body scattering experiments developed by S10. Guided by the fairly constant in time, N-independent behaviour of the MBHB hardening in the N-body runs, we conjectured that the N-body MBHB hardening rate is equivalent to that of a binary immersed in a field of unbound stars with density $\rho_{\rm inf}$ and velocity $\sigma_{\rm inf}$, equal to the density and velocity dispersion of the surrounding stellar distribution at the binary influence radius. We demonstrated the validity of this statement by showing that the dimensionless N-body hardening rate is comparable to that found in standard 3-body experiments, if normalized to $\rho_{\rm inf}$ and $\sigma_{\rm inf}$ (see table \ref{tab1}). We also showed that, when normalized to the same reference density and velocity dispersion, the MBHB evolution in the N-body runs and in the hybrid model matches fairly well, even in terms of eccentricity growth. The N-independent MBHB hardening seen in N-body runs, although fairly robust, has been tested up to $10^6$ particles only. \cite{2015MNRAS.446.3150V} developed a clever technique to extrapolate N-body results in the formal limit $N\rightarrow \infty$, finding hardening rates that are about a half of those found at $N=10^6$ \citep{2014arXiv1411.1762V}. However, these models are equilibrium triaxial systems, that might not capture non-relaxation and rotation effect present in merger remnants. Moreover, efficient diffusion in realistic systems due to the abundant presence of massive perturbers (globular clusters, giant molecular clouds) can easily bring the hardening rate back to the full loss cone regime \citep{2008ApJ...677..146P}.


Our findings have a variety of interesting applications and implications in several astrophysical contexts. Firstly, one can directly infer the lifetime of a putative MBHB residing in a gas poor galaxy solely based on the observed stellar density profile and on an estimate of the MBHB mass. In fact, those are the only ingredients needed to work out the the stellar density at the binary influence radius and to estimate the binary coalescence time through the hybrid model. This prescription can be reliably used in semianalytic models or large scale simulations of galaxy formation, where often the binary hardening phase is by-passed, assuming prompt coalescence. Secondly, the coalescence timescales we found for typical elliptical galaxies are longer than 1 Gyr, unless binaries are driven to considerably large eccentricities ($e>0.9$). Together with the fact that massive ellipticals had probably undergone at least one major merger since $z=1$ \citep[see, e.g.,][]{vandokkum05}, this implies that parsec scales MBHBs should be quite common in massive gas poor galaxies and their observational signatures might be accessible by future 30-40 meter optical facilities. Moreover, binary lifetimes in the Gyr range might lead to the frequent formation of MBH triplets following two subsequent mergers \citep{hoffman07}, which can have important consequences for the expected GW signal in the PTA band \citep{amaro10}, and we plan to investigate this point in future work. Thirdly, our results highlight the importance of knowing the eccentricity of the MBHB at formation. After the binary becomes hard, the eccentricity evolution can be reliably tracked using the hybrid model, but the value of the eccentricity when the binary first becomes bound cannot be predicted on the basis of simple arguments and it might well be stochastic (in fact in the K12 runs there is a large spread in the values of initial MBHB eccentricity). Highly eccentric binaries can coalesce much faster than their circular counterparts, having important implications on the MBHB population, on the occurrence of triplets, and on the nature of the GW signal in the PTA band, as discussed above. A large set of fairly short simulations following the MBHB only to right after it becomes bound would be extremely useful to complete the picture, and we are currently pursuing this. 

\section*{Acknowledgements}
The Authors thank Eugene Vasiliev for useful discussions, and the kind hospitality of the KITP at UCSB during the program {\it A Universe of Black Holes}, where this work was initiated. AS is supported by the University Research Fellow scheme of the Royal Society. FK thanks Kelly Holley-Bockelmann and Peter Berczik for hardware and software support of the simulations presented here which are performed on \textit{accre} GPU supported cluster at Vanderbilt University. 

\bibliographystyle{mn2e}
\bibliography{references}

\bsp

\label{lastpage}

\end{document}